\documentclass[final]{article}
\usepackage{graphicx,cite}
\usepackage{newproof}

\begin{document}

\title{A Library-Based Synthesis Methodology for \\ Reversible Logic}

\author{Mehdi Saeedi, Mehdi Sedighi, and Morteza Saheb Zamani \\
Quantum Design Automation Lab\\ Department of Computer Engineering and Information Technology\\ Amirkabir University of Technology\\ Tehran, Iran \\
\texttt{\{msaeedi, msedighi, szamani\}@aut.ac.ir}
 }
\date{}

\maketitle

\begin{abstract}
Synthesis of reversible logic has received significant attention in the recent years and many synthesis approaches for reversible circuits have been proposed so far. In this paper, a library-based synthesis methodology for reversible circuits is proposed where a reversible specification is considered as a permutation comprising a set of cycles. To this end, a pre-synthesis optimization step is introduced to construct a reversible specification from an irreversible function. In addition, a cycle-based representation model is presented to be used as an intermediate format in the proposed synthesis methodology. The selected intermediate format serves as a focal point for all potential representation models.

In order to synthesize a given function, a library containing seven building blocks is used where each building block is a cycle of length less than 6. To synthesize large cycles, we also propose a decomposition algorithm which produces all possible minimal and inequivalent factorizations for a given cycle of length greater than 5. All decompositions contain the maximum number of disjoint cycles. The generated decompositions are used in conjunction with a novel cycle assignment algorithm which is proposed based on the graph matching problem to select the best possible cycle pairs. Then, each pair is synthesized by using the available components of the library. The decomposition algorithm together with the cycle assignment method are considered as a binding method which selects a building block from the library for each cycle. Finally, a post-synthesis optimization step is introduced to optimize the synthesis results in terms of different costs.

To analyze the proposed methodology, various experiments are performed. Our analyses on the available reversible benchmark functions reveal that the proposed library-based synthesis methodology can produce low-cost circuits in some cases compared with the current approaches. The proposed methodology always converges and it typically synthesizes a give function fast. No garbage line is used for even permutations.
\end{abstract}

\newtheorem{thm}{Theorem}
\newtheorem{lem}{Lemma}
\newtheorem{example}{Example}
\newproof{pf}{Proof}

\section {Introduction}
An $n$-input, $n$-output, fully specified Boolean function is reversible if it maps each input pattern to a unique output pattern. A \emph{gate} is called reversible if it realizes a reversible function.
In 1961, Landauer proved that using conventional irreversible logic gates leads to a certain amount of energy dissipation per irreversible bit operation regardless of the underlying technology \cite{Landauer1961}. In 1973, Bennett stated that to avoid power dissipation in a circuit, it must be built from reversible gates \cite{Bennett1973}.

Energy consumption has become one of the most challenging problems in digital circuit design. To reduce power dissipation in CMOS circuits, numerous approaches have been proposed in the recent years which improve the non-ideal behavior of transistors and materials \cite{Rabaey2009}.
However, such methods cannot provide zero energy dissipation if irreversible bit operation is permitted \cite{Landauer1961}.

While heat generation due to the information loss in modern CMOS circuits seems to be small compared with the other parts of power dissipation, it has been shown that power dissipation resulted from information loss is at least 0.147 $W$ for a fully loaded Intel Itanium-2 processor \cite {Stinson2003}. In addition, heat removal will be more difficult with the increasing density of CMOS integrated circuits \cite{Zhirnov2003}. Currently, reversible computing has received considerable attention in particular in low-power CMOS design \cite{Schrom1998}.

Besides the power consumption problem of CMOS digital circuits, the unceasing miniaturization of integrated circuits is widely expected to end within the coming years \cite{ITRS}. This problem leads researchers to investigate new computational paradigms. Among them, quantum computing seems to be the most promising approach \cite{Maslov2007a}. Quantum gates are inherently reversible \cite {Nielsen2000}. Thus, reversible logic has also found great interest in the domain of quantum computation. As such, various Boolean reversible gates are used in different quantum algorithms \cite {Barenco1995}. While the advantages of quantum computing are not totally available without pure quantum gates, constructing efficient circuits with Boolean reversible gates is considered an important step towards realization of quantum systems \cite{Maslov2007a}, \cite{Shende2003}.

Boolean reversible circuit synthesis is defined as the ability to generate a reversible circuit from a given Boolean reversible specification. Synthesis of reversible logic differs from that of irreversible circuits because of various constraints imposed by the reversibility. For examples, loop and fanout are not allowed in reversible logic. Therefore, available irreversible synthesis approaches cannot be applied to synthesize reversible circuits as well.
To address this need, several synthesis algorithms for reversible functions have been proposed where both exact \cite {Hung2006,Grosse2009} and heuristic approaches \cite {Maslov2007a, Saeedi2007, Wille2009b, Gupta2006} have been applied.

Exact synthesis algorithms use methods such as Boolean satisfiability (SAT) \cite{Grosse2009} or symbolic reachability analysis \cite {Hung2006} to obtain optimal circuits for reversible specifications. More precisely, exact approaches define a set of equations to model the synthesis stage as a well-defined problem (e.g., SAT) first. Then, available solvers are applied to find at least one solution (i.e., a synthesized circuit) for the given specification. However, due to the exponential search space growth\footnote{Exact modelings are done based on the characterizations of the input specification such as the number of input lines and the number of required gates.} such approaches are useful to obtain optimal circuits for small specifications and they cannot be used to handle relatively large functions.

On the other hand, several heuristic methods have been proposed to find an efficient circuit for a given specification where the term `efficiency' can be defined according to various metrics \cite{Mohammadi2009}. Among the available metrics, `quantum cost' is widely accepted to be used in the synthesis stage. However, based on the selected target technology\footnote{Several different quantum computing technologies with different strengths and challenges have been developed so far. Examples are ion traps, quantum dots, linear optic and NMR. See \cite{Ross2008} for different quantum technologies.} the consideration of one specific metric may be more important than the others. For example, while the number of garbage lines can be ignored for Boolean CMOS reversible circuits, it is very important for quantum and Boolean reversible circuits used in quantum logic. Hence, approaches that use an arbitrary number of garbage lines (e.g., \cite{Wille2009b}) cannot be applied to quantum logic.

In \cite{Sasanian2009}, an NCT-based synthesis algorithm has been proposed that considers reversible functions as a set of cycles where each cycle was implemented by several reversible gates. By extending the results of \cite {Sasanian2009}, this paper proposes a library-based synthesis methodology for reversible circuits which uses the NCT gate library where binding and optimization methods along with a set of building blocks are introduced to be used in a unified library-based synthesis methodology. The rest of the paper is organized as follows. Basic concepts are introduced in Section \ref {basic_concept}. The synthesis algorithm of \cite {Sasanian2009} is described in Section \ref{sec:previous_work}. The proposed library-based synthesis methodology is introduced in Section \ref{method}. Experimental results are presented in Section \ref {exp_results}
and finally, Section \ref{conc} concludes the paper.

\section{Basic Concepts} \label {basic_concept}
\subsection{Reversible Logic}

Let $A$ be any set and define $f:A \rightarrow A$ as a one-to-one and onto transition function. The function $f$ is called a \emph{permutation function}, as applying $f$ to $A$ leads to a set with the same elements of $A$ and probably in a different order. If $A=\{1, 2, 3,\dots, m\}$, there exist two elements $a_i$ and $a_j$ belonging to $A$ such that $f(a_i)=a_j$. A \emph{k-cycle} with \emph{length} $k$ is denoted as $(a_1, a_2,\dots, a_k)$ which means that $f(a_1)=a_2$, $f(a_2)=a_3$, ..., and $f(a_k)=a_1$. A given $k$-cycle $(a_1, a_2, \dots, a_k)$ could be written in many different ways such as $(a_2, a_3, \dots, a_k, a_1)$. A cycle of length 2 is called \emph{transposition}.

Cycles $c_1$ and $c_2$ are called \emph{disjoint} if they have no common members, i.e., $\forall a_i \in c_1, a_i \notin c_2$. Any permutation can be written uniquely, except for the order, as a product of disjoint cycles. The unique cycle form of a permutation is called \emph{canonical cycle form} (\emph{CCF}) \cite {Sasanian2009}. If two cycles $c_1$ and $c_2$ are disjoint, they can commute, i.e., $c_1c_2= c_2c_1$. In addition, a cycle may be written in different ways as a product of transpositions, and using different numbers of transpositions. For example, the 3-cycle $(1,2,4)$ can be written as a product of two transpositions as $(1,2)(1,4)$. \label {transpositionexample}

A cycle (or a permutation) is called \emph{even} if it can be written as an even number of transpositions. A similar definition is introduced for an \emph{odd} cycle. Although there may be too many ways to \emph{decompose} a given cycle into a set of transpositions, the parity of the number of transpositions used remains the same, i.e., all resulted decompositions have the same even/odd number of transpositions. It can be verified that for a given even (odd) value of $k$, the resulted $k$-cycle can be written as an odd (even) number of transpositions. Hence, a $k$-cycle is odd (even) if $k$ is even (odd). Each reversible function can be considered as a permutation function.

A \emph{generalized Toffoli gate} C$^m$NOT ($x_1$, $x_2$, $\cdots$, $x_{m+1}$) passes the first $m$ lines unchanged. These lines are referred to \emph{control} lines. This gate flips the $(m+1)^{th}$ line (i.e., \emph{target}) if and only if the control lines are all one. Therefore, the generalized Toffoli gate works as follows: $x_i(out)=x_i (i<m+1), x_{m+1}(out)=x_1 x_2 \cdots x_m$ $\oplus$ $x_{m+1}$. For $m=0$ and $m=1$, the gates are called \emph{NOT} and \emph{CNOT}, respectively. For $m=2$, the gate is called C$^2$NOT or \emph{Toffoli}.

In addition to the C$^m$NOT gate, several other gates have been proposed previously \cite{Nielsen2000}. Among them, controlled-$V$ (controlled-$V^+$) changes the value on its target line using the transformation given by the matrix $V$ ($V^+$) if the control line has the value of 1.

\begin{equation}
V = \frac{{1 + i}}{2}\left[ {\begin{array}{*{20}c}
   1 & { - i}  \\
   { - i} & 1  \\
\end{array}} \right],V^ +   = \frac{{1 - i}}{2}\left[ {\begin{array}{*{20}c}
   1 & i  \\
   i & 1  \\
\end{array}} \right]
\end{equation}

To physically realize a synthesized circuit, all complex gate should be decomposed into a set of primitive gates. It has been shown that all one-qubit gates and a standard two-qubit gate, usually CNOT, can be used for such decomposition \cite{Nielsen2000,Shende2006}. In \cite{Smolin1996} all two-qubit quantum gates were used during the decomposition. The gates NOT, CNOT, controlled-$V$, and controlled-$V^+$ have been efficiently simulated in some quantum computer technologies \cite{Lee2006}. These gates were studied in the literature \cite{Nielsen2000} and are considered as \emph{elementary gates} for reversible Boolean functions \cite{Barenco1995}, \cite{Hung2004}. We used the same set of elementary gates throughout the paper. The number of elementary gates required for simulating a given gate is called \emph{quantum cost}. Inputs (outputs) that are not required in the specification of a reversible function are called \emph{constant} (\emph{garbage} or \emph{auxiliary}) bits.

\emph{Positive polarity Reed-Muller} (\emph{PPRM}) expansion can also be used to describe a reversible specification. PPRM expansion uses only un-complemented (or positive) variables and it can be derived from the EXOR-Sum-of-Products (\emph{ESOP}) description by replacing $a'$ with $a\oplus1$ for a complemented variable $a$. In addition, some algebraic manipulation of product terms may also be done to simplify the equations. The PPRM expansion of a function is canonical and is defined as:

\begin{equation}
\begin{array}{l}
 f(x_1 ,x_1 ,...,x_n ) = a_0  \oplus a_1 x_1  \oplus  \cdots  \oplus a_n x_n  \oplus a_{12} x_1 x_2  \oplus  \cdots  \\
 \,\,\,\,\,\,\,\,\,\,\,\,\,\,\,\,\,\,\,\,\,\,\,\,\,\,\,\,\,\,\,\,\,\,\, \oplus a_{n,n - 1} x_{n - 1} x_n  \oplus  \ldots  \oplus a_{12...n} x_1 x_2  \cdots x_n  \\
 \end{array}
\label{eq:pprm}
\end{equation}

\begin{figure*}[tb]
	\centering
		\includegraphics[scale=0.7]{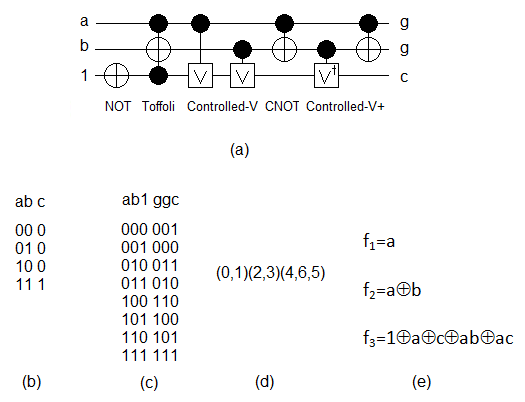}
		\caption{(a) A sample reversible circuit, (b) input specification in the truth table notation without constant and garbage lines and with constant and garbage lines (c), in CCF notation (d), and in PPRM notation (e)}
	\label{Fig:sampleCircuit}
\end{figure*}

A sample reversible circuit which includes one constant line with the initial value 1 and two garbage lines (i.e., shown by symbol \emph{g}) is depicted in Figure \ref{Fig:sampleCircuit}. The input specification in different notations are also illustrated in this figure.

It has been shown that for $n\geq5$ and $m \in \left\{3,4, \cdots \left\lceil n/2 \right\rceil\right\}$, a C$^m$NOT gate can be simulated by 12$m$-22 elementary gates. In addition, for $n \geq 7$, a C$^{n-2}$NOT gate can be simulated by 24$n$-88 elementary gates with no auxiliary bits \cite{Maslov2008}. On the other hand, a C$^{n-1}$NOT gate can be simulated with an exponential cost 2$^n$-3 if no garbage line is available \cite{Barenco1995}. To avoid the exponential size and the need for a large number of elementary gates, several researchers used an extra garbage line for an efficient simulation of C$^{n-1}$NOT gate \cite{Maslov2007a}. Generally, the number of available bits is very restricted in today's reversible and quantum implementations \cite{Negrevergne2006}. Therefore, for two circuits with equal linear costs, the one without garbage line is preferred.

\subsection{Cycle Factorization}
A reversible specification can be considered as a permutation function which includes a set of cycles of various lengths. On the other hand, a given cycle of length greater than two can be \emph{factorized} into several cycles of smaller lengths. Let $\sigma_1, \cdots, \sigma_m$ be a factorization of the cycle $(a_1, a_2, $\ldots$, a_n$) into a product of smaller cycles. We say the factorization is of \textit{type} $\alpha=(\alpha_2, \dots, \alpha_k)$ if among $\sigma_j (1 \leq j \leq m)$ there are exactly $\alpha_2$ 2-cycles, $\alpha_3$ 3-cycles and so on. Let us define:

\begin{equation}
\langle \alpha \rangle=\Sigma_{j\geq 2}(j-1) \alpha_j
\label {eq1}
\end{equation}
where $\alpha$ satisfies $\langle \alpha \rangle \geq n-1$. For the case of equality, the factorization is called \emph{minimal}. Two cycle factorizations are called \emph{equivalent} if one can be obtained from the other by repeatedly exchanging adjacent factors that are disjoint.

\begin{example}\label{ex:factor}
Consider a given cycle $\pi$=($a$, $b$, $c$, $d$, $e$) of length $n=5$. It can be verified that $\pi$ can be factorized into ($a$, $b$) ($a$, $c$) ($a$, $d$, $e$) with the cycle type (2, 1). Note that cycles are applied from left to right. For this factorization, we have $\langle \alpha \rangle$=$1 \times 2 + 2 \times 1=4$. Since $\langle \alpha \rangle = n-1 $ this factorization is minimal.
\end{example}

Cycle factorization has a rich history in combinatorial problems \cite{berkolaiko2008,Hurwitz1891,Eidswick1989}. In particular, a significant effort has been directed to count the number of $k$-cycle factorizations. The case $k = 2$ (transposition factors) is known as the \textit{Hurwitz problem} \cite{Hurwitz1891}. The following formula gives the number of 2-cycle factorizations of any permutation of cycle type $(\alpha_1, \dots, \alpha_m)$:

\begin{equation}
n^{(m-3)} (n+m-2)! \Pi_{i=1}^m \frac {\alpha_i^{\alpha_i+1}} {\alpha_i!}
\end{equation}

It has been proved that the number of inequivalent 2-cycle factorizations of the cycle $(1, 2, \dots n)$ is the generalized Catalan number \cite{Eidswick1989}:

\begin{equation}
\frac {1} {2n-1} (^{3n-3}_{n-1})
\end{equation}

The following theorems examine the number of cycle factorizations for general cases. In this paper, cycle factorization is used to extract library elements from a given reversible specification as discussed in Section \ref {method} in detail.

\begin {thm} \label {cycle_factorization}
Let $i = (i_2, i_3, \dots)$ be a sequence of nonnegative integers and set $r = r(i) = i_2 + i_3 + \dots $. Then, the number of cycle factorizations of $(1,2, \dots, n)$ with cycle index $i$ is
\begin{equation}
H_{[n]}(i)=\frac{n^{r-1} r!} {\Pi_{k \geq 2} i_k !}
\end{equation}
in the case that $n+r-1=\Sigma_{k \geq 2} k i_k$, and zero otherwise.
\end {thm}

\begin {thm} (from \cite {Springer1996}) \label {inequivalent_cycle_factorization}
Let $i = (i_2, i_3, \dots)$ be a sequence of nonnegative integers, not all zero, $r = r(i) = i_2 + i_3 + \dots$. Then, the number of inequivalent cycle factorizations of  $(1,2, \dots, n)$ with cycle index $i$ is
\begin{equation}
\tilde{H}_{[n]}(i) = \frac {(2n+r-2)!} {(2n-1)! {\Pi_{k \geq 2} i_k !}}
\end{equation}
if $n+r-1=\Sigma_{k \geq 2} k i_k$, and zero otherwise.
\end {thm}

\subsection {Graph Matching}
In order to select library elements in the proposed library-based synthesis methodology (Section \ref{sec:CA}), an available graph perfect matching algorithm is applied.
Given a graph $G = (V,E)$, a \textit{matching} $M$ in $G$ is a set of pairwise non-adjacent edges; that is, no two edges share a common vertex. A vertex is \textit{matched} if it is incident to an edge in the matching. Otherwise the vertex is \textit{unmatched}.
A \textit{maximum matching} is a matching that contains the largest possible number of edges.
There may be many maximum matchings. 

A \textit{perfect matching} is a matching which matches all vertices of the graph. That is, every vertex of the graph is incident to exactly one edge of the matching.
In a weighted bipartite graph, each edge has an associated value. A \textit{minimum weighted bipartite matching} is defined as a perfect matching where the sum of the values of the edges in the matching has a minimal value. If the graph is not complete bipartite, missing edges are inserted with value zero.

\section {Previous Work} \label {sec:previous_work}
Several authors discussed the requirements of a design methodology for reversible and quantum circuits. In \cite{Svore2006}, a computer-aided design flow for quantum computation was presented that transforms a high-level language program into a technology-specific implementation. In addition, the languages and transformations needed to represent and optimize a quantum algorithm in the proposed design flow were discussed. The authors of \cite{Udrescu2004} introduced an HDL-based simulation methodology for quantum circuits where the HDL feature of describing a circuit with both structural and functional architectures was employed in the proposed methodology. In \cite{Balensiefer2005}, the authors proposed an instruction set architecture and several tools such as compiler, device scheduler and simulator for ion trap based quantum computers. A computer-aided design flow for quantum circuits was proposed in \cite{Whitney2007} which includes automatic layout and control logic extraction. In addition, several heuristics for the placement and routing of quantum circuits in ion trap technology were presented in \cite{Whitney2007}. In the following paragraphs, those papers published for the synthesis of reversible circuits are discussed.

The synthesis of reversible circuits composed of generalized Toffoli gates has been studied extensively \cite{Maslov2007a,Saeedi2007, Grosse2008, Wille2009b, Gupta2006}. Since the cost of a generalized Toffoli gate in terms of the physical implementation is high, to realize a complex generalized Toffoli gate it should be decomposed into some elementary gates \cite{Maslov2008}. Although this approach was adopted more in the previous years, a direct synthesis method that uses simple elementary gates could behave more efficiently. To this end, a few papers \cite{Shende2003, Prasad2006, Sasanian2009} were published in recent years which used \emph{NCT gate library} containing simple low-cost NOT (N), CNOT (C) and Toffoli (T) gates.

The authors of \cite{Shende2003} proposed an NCT-based synthesis method which applies N, T, C and T gates in order (i.e., the T$|$C$|$T$|$N method) to synthesize a given permutation. In the first C$|$T$|$N part, the terms 0 and 2$^i$ of a given reversible function are positioned at their right locations while the last Toffoli network places the other truth table terms in their right positions.

In \cite{Shende2003}, for the last Toffoli part, a given $k$-cycle is decomposed into a set of transpositions. Subsequently, each pair of disjoint transpositions ($a$, $b$) ($c$, $d$), is implemented by a circuit (i.e., the $\pi$ circuit) that maps $a$, $b$, $c$ and $d$ to $2^n-4$, $2^n-3$, $2^n-2$ and $2^n-1$, respectively where $n$ is the number of bit in the function specification. Then, the permutation ($2^n-4$, $2^n-3$) ($2^n-2$, $2^n-1$) is implemented by a circuit called $\kappa_0$. Finally, the reverse $\pi$ circuit, i.e., $\pi^{-1}$, is applied to transform $2^n-4$, $2^n-3$, $2^n-2$ and $2^n-1$ into $a$, $b$, $c$ and $d$, respectively. It can be verified that the $\pi \kappa_0 \pi^{-1}$ circuit implements the permutation ($a$, $b$) ($c$, $d$). An extension of \cite{Shende2003} was suggested in \cite{Prasad2006} which produces better quantum cost by applying the unit-cost NOT and CNOT gates instead of using Toffoli gates with cost 5 in many situations.

In our previous work \cite{Sasanian2009}, a cycle-based synthesis algorithm was proposed based on the results of \cite{Prasad2006} where cycles of lengths less than 4 are synthesized directly. More exactly, in \cite{Sasanian2009} a set of synthesis algorithms were proposed to synthesize a pair of 2-cycles, a single 3-cycle, and a pair of 3-cycles. Each cycle is called a \emph{building block} or an \emph{elementary cycle}. In order to improve the synthesis cost, the authors extended the building blocks to include a single 4-cycle followed by a single 4-cycle or a single 2-cycle, a single 5-cycle and a pair of 5-cycles. In addition, we used NOT and CNOT gates instead of Toffoli in many situations.

\begin{figure}[tb]
	\centering
		\includegraphics[scale=0.55]{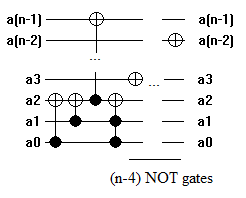}
		\caption{The $\pi_2$ circuit for the (2,2) synthesis algorithm \cite {Sasanian2009}.}
	\label{Fig:picircuit}
\end{figure}

\begin{figure}[tb]
	\centering
		\includegraphics[scale=0.55]{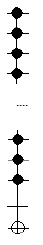}
		\caption{The $\kappa_0$ circuit \cite {Sasanian2009}}
	\label{Fig:k0circuit}
\end{figure}

\begin{figure*}[tb]
	\centering
		\includegraphics[scale=0.5]{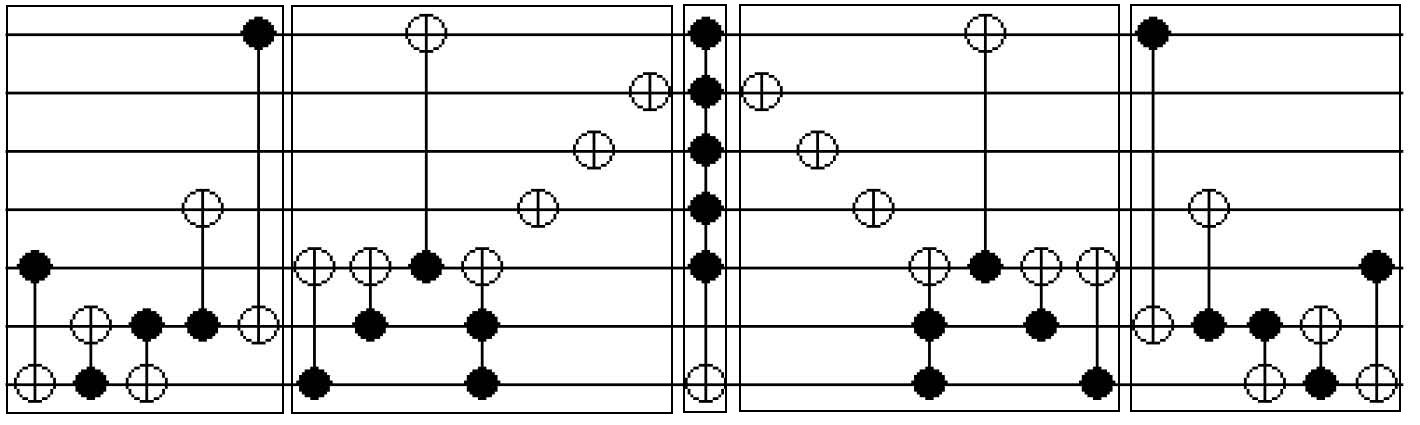}
		\caption{A sample circuit synthesized by the method of \cite {Sasanian2009}.}
	\label{Fig:sample22}
\end{figure*}

\begin{example}\label{ex:22}
Assume that the pair of 2-cycles $(5, 3)$ $(9, 67)$ should be implemented. To this end, the term 5 is transformed to 4 by a CNOT gate (gate \#1 in Fig. \ref{Fig:sample22}) which has no effect on other terms. Similarly, 3 is transformed to 1 by a CNOT gate (gate \#2 in Fig. \ref{Fig:sample22}) which changes the term 9 to 11 and 67 to 65. Then, 11 is transformed to 2 by two CNOT gates (gate \#3, gate \#4 in Fig. \ref{Fig:sample22}) with no effect on other terms. Finally, 65 is transformed to 67 by a CNOT gate (gate \#5 in Fig. \ref{Fig:sample22}). Then, a pre-designed circuit, such as the one shown in Figure \ref {Fig:picircuit} ($\pi_2$), is applied followed by the circuit shown in Figure \ref {Fig:k0circuit} ($\kappa_0$). Afterwards, the gates applied before the $\kappa_0$ circuit are applied in the reverse order. Fig. \ref{Fig:sample22} illustrates the complete circuit.
\end {example}


On the other hand, to synthesize a given \emph{large} cycle of length $k$ ($k>3$) the authors used one possible decomposition to extract the suggested building blocks (i.e., cycles) from the input specification (i.e., permutation) that leads to a set of cycles of lengths 3 and probably a cycle of length less than 3. Since we used an extended set of building blocks here, the decomposition algorithm was modified to detach 5-cycles. Therefore, the results of the decomposition algorithm is a set of cycles of lengths 5 and probably a cycle of length less than 5. As the synthesis of a cycle  pair is more efficient than the synthesis of two single cycles by using the method of \cite{Sasanian2009}, cycle pairs are explored during the synthesis as discussed in the following sections in details.

\begin{example}\label{ex:dcm}
Consider a given permutation $\pi$=(3, 5, 6, 7, 9, 10, 11, 12, 13, 14, 15, 17, 18, 19, 20, 21) (22, 23, 24, 25, 26, 27) (28, 29) (30, 31). It can be verified that applying the decomposition algorithm of \cite {Sasanian2009} for detaching all 5-cycles leads to $\pi$=(3, 5, 6, 7, 9) (10, 11, 12, 13, 14) (15, 17, 18, 19, 20) (22, 23, 24, 25, 26) (21, 3, 10, 15) (22, 27) (28, 29)(30, 31).	
\end{example}
\label{libsize}
Let $d_{r_1,r_2,\dots,r_k}(n,k)$ be the number of permutations with exactly $k$ cycles of length $r_1,r_2,\dots,r_k$ for a set of $n$ distinct numbers. The falling factorial $(n)_k$ is defined as $n(n-1)(n-2)\dots(n-k+1)$. The size of each building block can be determined as $d_{2,2}(n,2)=(n)_4$, $d_{3}(n,1)=(n)_3$, $d_{3,3}(n,2)=(n)_6$, $d_{4,2}(n,2)=(n)_6$, $d_{5}(n,1)=(n)_5$, $d_{5,5}(n,2)=(n)_{10}$. To prove, consider a pair of two cycles $(a,b)(c,d)$. For the first element $a$, all $n$ elements can be selected. For the next element $b$, $n-1$ elements can be selected and so on. \\
In the next section, we propose a library-based synthesis methodology for reversible circuits based on the results of \cite {Sasanian2009}.

\section {The Proposed Synthesis Methodology} \label{method}
\begin{figure}[!tb]
	\centering
		\includegraphics[scale=0.5]{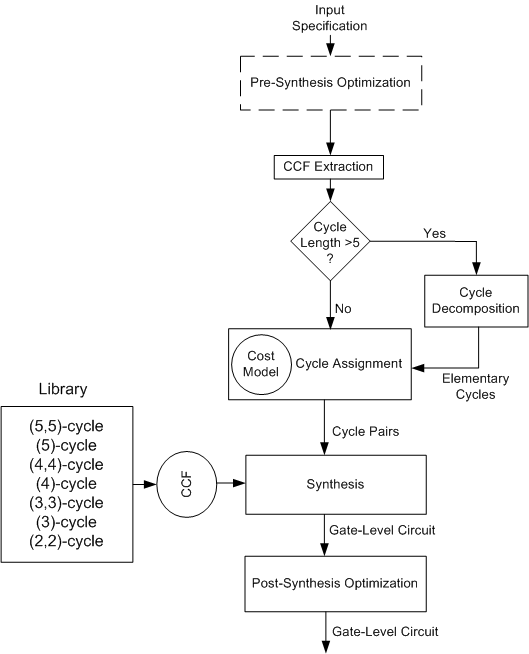}
		\caption{Proposed Synthesis Methodology}
	\label{Fig:flow}
\end{figure}

The proposed synthesis methodology is shown in Figure \ref{Fig:flow}. In order to synthesize a given input specification, a pre-synthesis optimization is applied on the given function to improve it with respect to some metrics (Section \ref{sec:pre_opt}). Subsequently, a CCF representation is extracted from the prepared input specification (Section \ref{sec:IF}) and then is gradually mapped into a reversible circuit. To this end, if cycle length is greater than 5, we apply a cycle decomposition algorithm (Section \ref{sec:DCM}) to construct elementary cycles. Next, a cycle assignment method (Section \ref {sec:CA}) is applied to construct cycle pairs based on the well-known graph matching problem. Then, each pair is synthesized by applying the method of \cite {Sasanian2009} and finally a post-synthesis optimization is applied to improve the circuit cost (Section \ref{sec:post_opt}).

\subsection{Pre-Synthesis Optimization} \label {sec:pre_opt}
As discussed in Section \ref{basic_concept}, an $n$-input, $n$-output, fully specified Boolean function is reversible if it maps each input pattern to a unique output pattern. Hence, a reversible specification must have `the same number of inputs and outputs' with `unique assignments'\footnote {Recall the truth table notation of reversible circuits. For a reversible circuit with $n$ inputs and $n$ outputs (i.e., a circuit of size $n$), a truth table of size $n \times 2^n$ is required where the values of outputs are uniquely selected from 0 to $2^n-1$ probably with a different order. The goal of a synthesis algorithm is to put outputs at their right locations (i.e., 0 to $2^n-1$ sequentially) by applying a set of reversible gates.}. For example, reconsider the circuit shown in Figure \ref{Fig:sampleCircuit}-(a) which contains one constant input and two garbage outputs. As illustrated in Figure \ref{Fig:sampleCircuit}-(b), the initial function specification (without constant and garbage lines) does not have the characteristics of a reversible specification. However, after the insertion of constant and garbage lines and unique output assignments (see Figure \ref{Fig:sampleCircuit}-(c)) a reversible specification of size 3 is resulted.

Since the values of constant and garbage lines and their locations with respect to other lines are not in the initial specification of an irreversible function, such parameters can be manipulated by a synthesis tool to improve the final cost. Hence, the values of constant and garbage lines are called \emph{don't cares} (\emph{DC}). The goal of the pre-synthesis optimization is to assign appropriate values to DC lines (\emph{DC assignment}) and place them at proper locations (\emph{constant and garbage assignment}) to improve the cost. Such optimizations are mandatory for irreversible specifications and can be ignored if completely specified functions are addressed as done in this paper.

It is worth noting that some DC assignment algorithms have been proposed recently \cite{Grosse2008}. However, as the efficiency of such assignments depends on the characteristics of a synthesis algorithm, it is not possible to use a well-developed pre-synthesis optimization algorithm for all synthesis approach\footnote{Some authors reordered the locations of output lines of a given fully specified specification to improve its final cost \cite{Wille2009a}. While reordering circuit lines changes the original function specification, it may be acceptable for some applications. This approach can also be considered as a pre-synthesis optimization method.}.

\subsection{Intermediate Format} \label {sec:IF}

Different synthesis algorithms used different representations for their input specifications. Among the available models, truth table \cite{Maslov2007a, Saeedi2007,Maslov2008} and PPRM expansions \cite{Gupta2006, Saeedi2007a} have been widely used. The selected model works as an \emph{intermediate format} (\emph{IF}) for the respective synthesis algorithm and is placed between two levels of abstraction (i.e., input specification and gate-level circuit). In this work, CCF representation has been selected as an IF as discussed below\footnote{CCF has been used to describe the input specification in \cite{Sasanian2009, Shende2003} to some extent. However, using CCF as an intermediate format is introduced here for the first time.}.

Compared with the truth table model, CCF removes fixed rows\footnote{A truth-table row is called \emph{fixed} if it is mapped into itself.} of a given truth table and hence are very efficient for large functions particularly those that map many input combinations into themselves. Recall that
a synthesis algorithm returns the changed rows to their right positions. To this end, a set of reversible gates is applied where fewer gates lead to fewer cost, generally. Fixed rows can be removed to save memory if the synthesis algorithm does not use them directly. In \cite{Maslov2007a}, the authors reported that the applicability of their synthesis algorithm was limited due to the memory constraint occurred during the representation of large input specifications in the truth table format.

Moreover, while some truth table-based approaches like the one introduced in \cite{Maslov2007a} considered both input-to-output and output-to-input transformations at the same time (namely \emph{bidirectional method}), the mentioned transformations have equal CCF representations. Therefore, there is no need to consider both transformations at the synthesis step concurrently. Hence, lower complexities should be handled by the synthesis method.

On the other hand, for a given $n$-input,$n$-output reversible function, $n$ PPRM expansions can be extracted which remove explicit values of truth table rows. Of course, the truth table rows can be recovered from the PPRM expansions with further processing cost. While PPRM notation received attractions in some synthesis algorithms, it cannot be used in the proposed method since explicit row values are needed in this paper. Altogether, CCF benefits from compact notation of PPRM expansions as well as explicit values of truth table representation. Therefore, CCF is used as the selected IF in the proposed synthesis methodology.

Having an input specification in the CCF format, the next step is to synthesize it according to \cite {Sasanian2009} where small cycles are synthesized by the suggested building blocks directly. Table \ref{dis_bench} shows the average distribution of cycle lengths for the benchmark functions \cite{Grosse2008}. As shown in this table, more than 60\% of cycle lengths are greater than 5. Therefore, many cycles should be decomposed into the proposed set of cycles and hence, cycle decomposition can affect the synthesis costs considerably. In the following, the effects of cycle decomposition on the synthesis results are examined.

\begin{table}[!tb]
\caption{Average distribution of cycle lengths for available benchmark functions}
\label{dis_bench}
\centering
\begin{tabular}{|c|c|c|c|c|c|}
\hline
2-cycle & 3-cycle & 4-cycle & 5-cycle & $\geq$ 6-cycle \\
\hline
\hline
18.95\%	& 6.57\%	& 8.79\%	& 2.87\%	& 62.82\% \\
\hline
\end{tabular}
\end{table}

\subsection{Cycle Decomposition} \label{sec:DCM}
Since each decomposed cycle should be synthesized by a set of reversible gates, reducing the number of decomposed cycles is preferred in \cite{Sasanian2009} to reduce final cost. Moreover, the cycles produced by decomposing disjoint cycles are disjoint too. Hence, they can commute to find the best possible selection of cycle pairs for having lower synthesis cost. Altogether, each large cycle should be decomposed so that the minimal number of inequivalent 5-cycles are generated and the number of disjoint 5-cycles are maximized. For a given cycle $\pi$ of length $n$, $N_5(n)$ is used as the minimum number of inequivalent decomposed 5-cycles.

More precisely, to decompose a given large cycle of length $n$ into a set of 5-cycles, we impose the following conditions:
\begin{itemize}
\item All decomposed cycles should be of length 5 except at most one cycle which is of length less than 5.
\item Cycle factorization should be minimal.
\item Inequivalent cycle factorization is considered.
\item Maximum number of disjoint cycles should be produced.
\end{itemize}

The first three conditions can be addressed by using the results of Theorem \ref {inequivalent_cycle_factorization} for $i_j=1$ where $j$ is equal to 2, 3 or 4 and $i_5=N_5(n)$. To address the last condition some modifications are required.

\begin{lem}
Consider a cycle $\pi$ of length $n$. The maximum number of disjoint cycles resulted from an inequivalent 5-cycle factorization is $\lfloor n/5 \rfloor$.
\end{lem}

\begin{pf}
Since there are $n$ distinct elements in $\pi$ and each 5-cycle has five distinct elements, at most $\lfloor n/5 \rfloor$ disjoint 5-cycles can be resulted.\qed
\end{pf}

For a cycle $\pi$ of length $n$, assume that all disjoint 5-cycles are detached. According to the minimal factorization together with Equation (\ref{eq1}), we have $4 \times \lfloor n/5 \rfloor + (L - 1) = n - 1$ where $L$ is the length of the resulted non-disjoint cycle after detaching all disjoint 5-cycles (denoted as $\acute{\pi}$ in the following). It can be verified that $L$ is equal to $n-4\times \lfloor n/5 \rfloor$. Note that $\acute{\pi}$ includes at most four elements of $\pi$ which does not belong to the detached 5-cycles. In addition, it has $\lfloor n/5 \rfloor$ elements of $\pi$ each of which belongs to exactly one disjoint cycle inserted to recover the original cycle $\pi$ from the set of disjoint 5-cycles. Considering the minimal length of $\acute{\pi}$, there is exactly one element in $\acute{\pi}$ for each disjoint 5-cycle.
\begin{lem}
Consider a cycle $\pi$ of length $n$. The minimum number of decomposed 5-cycles, $N_5(n)$, resulted from an inequivalent 5-cycle factorization is
\begin{equation}
N_5 (\pi ) = \left\{ \begin{array}{l}
 \frac{{n - 4}}{4}\,\,\,\,\,\,\,\,\,\,\,\,\,\,\,\,\,\,\,\,\,\,\,\,n\mathop  \equiv \limits^4 0 \\
 \frac{{n - \bmod (n,4)}}{4}\,\,\,\,\,\,otherwise \\
 \end{array} \right.
\end{equation}
\end{lem}

\begin{pf}
Based on the definition of inequivalent 5-cycle factorization and Equation (\ref {eq1}), we have $\langle \alpha \rangle=(j-1) \times \alpha_j + 4 \times N_5(n)$ if $j$=2, 3 or 4 and
\[
\alpha _j  = \left\{ \begin{array}{l}
 1\,\,\,\,n\mathop  \equiv \limits^4 0,2,\,3 \\
 0\,\,\,n\mathop  \equiv \limits^4 1 \\
 \end{array} \right.
\]
Considering the definition of minimal factorization $\langle \alpha \rangle=n-1$, and by doing some arithmetic manipulations the lemma is proved.\qed
\end{pf}

In order to have \emph{both} the minimum number of decomposed cycles and the maximum number of disjoint cycles for a given cycle $\pi$, the order of elements in each disjoint cycle should be the same as the original cycle $\pi$; otherwise some extra cycles should be inserted to construct the given permutation. Consider the following example for more detail:

\begin{example}\label{ex:minfactor}
Consider a cycle $\pi$ = (1, 2, 3, 4, 5, 6, 7, 8, 9, 10, 11, 12, 13, 14, 15, 16, 17, 18, 19) of length $n=19$. Accordingly, $N_5(n)=4$ and $\lfloor n/5 \rfloor=3$. Therefore, a total of four decomposed 5-cycles exist three of which are disjoint. The decomposition $\pi$ = (1,2,3,4,5) (6,7,8,9,10) (11,12,13,14,15) (16,17,18,19,1) (6,11,16) meets the conditions. On the other hand, if the order of elements in disjoint cycles are modified, several extra cycles should be included. As an example, the decomposition $\pi$ = (1,2,4,5,6) (3,7,8,9,10) (11,12,13,14,15) (16,17,18,19,1) (3,11,16) (4,3,7) is not minimal.
\end{example}

According to the above discussion, the elements of each 5-cycle should have exactly the same ordering of the original large cycle $\pi$. Now, let us examine the elements of $\acute{\pi}$. As explained, it may contain at most four elements which do not belong to any disjoint 5-cycle. Consider $a_k \in \pi$ where $a_k$ does not belong to any detached disjoint 5-cycle. There are three cases regarding the element $a_k$ as follows:
\begin{itemize}
	\item Three successive elements $a_{k-1}$, $a_k$, and $a_{k+1}$ belong to $\acute{\pi}$
	\item Two successive elements $a_{k-1}$, $a_k$ or $a_k$, $a_{k+1}$ belong to $\acute{\pi}$
	\item Only $a_k$ belongs to $\acute{\pi}$
\end{itemize}
It can be verified that the predecessor ($a_{k-1}$) and the successor ($a_{k+1}$) of $a_k$ for the first case were placed at right locations. On the other hand, for the second and the third cases, some extra cycles should be inserted to fix the locations of the predecessor or the successor (or both) elements. Therefore, to have the minimum number of decomposed 5-cycles, the ordering of those elements which do not belong to any disjoint cycle should be the same as the original large cycle.

\begin{thm}\label{thm:3}
Consider a cycle $\pi=(a_1,a_2...,a_n)$ of length $n>5$ which should be decomposed into minimum number of inequivalent 5-cycles, $N_5(n)$, where the number of disjoint 5-cycles should be maximized (i.e., $\lfloor n/5 \rfloor$). Assume that $L^{(0)}=n$ and $L^{(i)}=L^{(i-1)}-4\times \lfloor L^{(i-1)}/5 \rfloor$. Then, there are $N_{DCM}(n)=L^{(0)} \times L^{(1)} \times \dots L^{(i)}$ ways for such factorizations where $L^{(i+1)}<5$.
\end{thm}

\begin{pf}
To have minimum number of inequivalent 5-cycles and maximum $\lfloor n/5 \rfloor$ disjoint 5-cycles, the ordering of elements in each disjoint 5-cycle should be the same as the ordering of $\pi$. Moreover, for those elements which do not belong to any disjoint 5-cycles, the same ordering of $\pi$ should be used in $\acute{\pi}$. Therefore, the sequence of all elements should be saved. Since we have $\pi=(a_1,a_2...,a_n)=(a_2,a_3...,a_n,a_1)=...=(a_n,a_1,...a_{n-1})$, there are $L^{(0)}=n$ ways of such decomposition. After detaching all disjoint 5-cycles, a non-disjoint cycle $\acute{\pi}$ of length $L^{(1)}=n-4\times \lfloor n/5 \rfloor$ will be resulted which can be decomposed into a set of 5-cycles in $L^{(1)}$ ways. This process can be continued until a non-disjoint cycle of length less than 5 is produced. Considering all ways of decompositions leads to the theorem.\qed
\end{pf}
For a given cycle $pi$ of length $n$, a recursive procedure can be applied in relation to the proof of Theorem \ref{thm:3} to extract all decompositions. Figure \ref{Fig:pseudo_code} illustrates a pseudo code.

\begin{figure}[!tb]
	\centering
        \begin{small}
        \begin{verbatim}
        ExtractDecompositions(pi){
          if n>5 {//at least one cycle can be extracted
            for i in 0 to n{
              FirstDecompositionResults =
                extract all disjoint cycles starting from the ith element of pi
              newCycle = construct a non-disjoint cycle pi'
              SecondDecompositionResults = CycleExtraction(newCycle)
              for j in 0 to length of SecondDecompositionResults{
                AllPossibleDecomposition[j] =//merge the results
                  {FirstDecompositionResults, SecondDecompositionResults[i]}
              }
            }
          }else{
            return pi
          }
          return AllPossibleDecomposition
        }
        \end{verbatim}
        \end{small}
	\caption {Extracting all possible decompositions} \label{Fig:pseudo_code}
\end{figure}

\begin{example}\label{ex:alldcm}
Consider a cycle $(1,2,3,...,18)$ of length $n=18$. This cycle can be decomposed into $\lfloor 18/5 \rfloor=3$ disjoint cycles in $L^{(1)}=18$ ways. After detaching all disjoint cycles, a non-disjoint cycle of length $L^{(2)}=18 - 4 * \lfloor 18/5 \rfloor=6$ is produced which can be decomposed into $\lfloor 6/5 \rfloor=1$ 5-cycle in $L^{(2)}=6$ different ways. Hence, $18 \times 6$ different decompositions are generated. The following items list four possible decompositions:
\begin{itemize}
\item $(1,2,3,4,5)(6,7,8,9,10)(11,12,13,14,15)(16,17,18,1,6)(11,16)$
\item $(1,2,3,4,5)(6,7,8,9,10)(11,12,13,14,15)(17,18,1,6,11)(16,17)$
\item $(1,2,3,4,5)(6,7,8,9,10)(11,12,13,14,15)(18,1,6,11,16)(17,18)$
\item $(18,1,2,3,4)(5,6,7,8,9)(10,11,12,13,14)(15,16,17,18,5)(15,10)$
\end{itemize}
\end{example}
There are many ways of decomposing a given large cycle into a set of cycles of length less than 6 with minimum number of decomposed cycles and maximum number of disjoint cycles. In the next subsection, the process of selecting cycle pairs is evaluated.

\subsection {Cycle Assignment} \label{sec:CA}

\begin{figure}[!tb]
	\centering
		\includegraphics[scale=0.4]{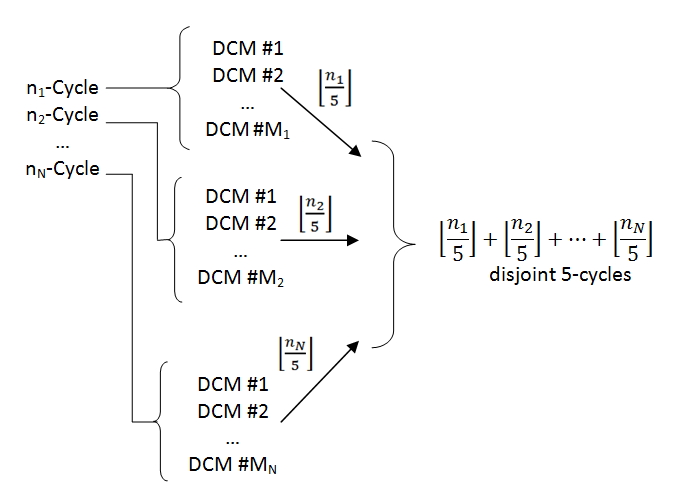}
		\caption{The results of cycle decomposition step}
	\label{Fig:CA1}
\end{figure}

For a given cycle $\pi$ of length $n$, $N_{DCM}(n)$ different decompositions are possible where each decomposition includes $N_5(n)$ decomposed 5-cycles with $\lfloor n/5 \rfloor$ disjoint 5-cycles. Non-disjoint cycles cannot be arbitrarily moved. Figure \ref {Fig:CA1} illustrates the result of cycle decomposition step. In this figure, an input specification with $N$ cycles are shown where the $i^{th}$ cycle was decomposed into $\lfloor n_i/5 \rfloor$ 5-cycles in $M_i$ different ways ($n_i > 5$ and $M_i = N_{DCM}(n_i))$ denoted as DCM \#1, $\cdots$, DCM \#$M_i$. Now, one can select one of the available decompositions for each input cycle to construct a set of elementary cycles of size $\lfloor n_1/5 \rfloor + \lfloor n_2/5 \rfloor+ \cdots + \lfloor n_N/5 \rfloor$. Next, cycle pairs should be assigned to be used by the synthesis algorithm as follows.

In order to find cycle pairs, we model the cycle assignment step as a graph perfect matching problem. For a set with $N$ elementary cycles, $N \times (N-1)/2$ cycle pairs can be determined where each pair can be synthesized with a specific quantum cost. Since each cycle pair can be considered as a valid cycle assignment, we first synthesize each cycle pair using the method of \cite {Sasanian2009}. Then, a weighted graph is constructed with $N$ nodes and $N \times (N-1)/2$ edges. The actual synthesis quantum cost for each cycle pair is used as the weight of the edge between the respective nodes. Next, a graph perfect matching algorithm is applied to find the best possible matching with the minimum cost. Therefore, cycle assignments which produce lowest total cost are found.

\begin{figure}[!tb]
	\centering
		\includegraphics[scale=0.5]{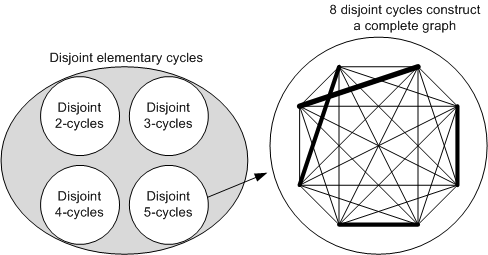}
		\caption{Cycle assignment. Different nodes represent different disjoint cycles. A connected edge between two nodes denotes the probability of synthesizing the cycles as a pair. Each edge contains a weight which is the synthesis cost of the involved cycles.}
	\label{Fig:CA2}
\end{figure}

Figure \ref{Fig:CA2} illustrates the cycle assignment problem for the generated disjoint 5-cycles. As can be seen in this figure, there are 8 disjoint 5-cycles which construct a complete graph on 8 nodes. A possible cycle assignment is shown by solid edges. It is worth noting that since all cycles of a given input specification are disjoint, the resulted set of 2-cycles contains only disjoint cycles. Therefore, it is possible to apply cycle assignment step to the elementary 2-cycles too. Similarly, this process can be repeated for all 3-cycles and 4-cycles.
\begin{example} \label{ex:ca}
Consider a given input specification with two cycles $\pi_1$ and $\pi_2$ of lengths 18 and 13, respectively. It can be verified that $N_{DCM}(18)=108$ and $N_{DCM}(13)=13$. In addition, the decomposition of $\pi_1$ and $\pi_2$ leads to $\lfloor 18/5 \rfloor=3$ and $\lfloor 13/5 \rfloor=2$ disjoint cycles. Therefore, a set of five disjoint cycles will be resulted. Now, a complete weighted graph with 5 nodes and 10 edges is constructed where nodes represent cycles and edges represent the probability of synthesizing the connected cycles as a pair. Edge weights are the actual synthesis costs. After running the perfect matching algorithm, two cycle pairs are selected to be synthesized with each other and the remaining cycle is synthesized alone.
\end{example}

In addition to the effect of cycle assignment on the synthesis cost, the order of elements in each cycle affects the synthesis result. More precisely, consider two disjoint 5-cycles $\pi_1^{(1)}=(a_1,a_2,a_3,a_4,a_5)$ and $\pi_2^{(1)}=(a_6,a_7,a_8,a_9,a_{10})$ where $a_i\neq a_j$ if $i\neq j, 1\leq i,j\leq 10$. It can be seen that these cycles can be written, for example, as $\pi_1^{(2)}=(a_4,a_5,a_1,a_2,a_3)$ and $\pi_2^{(2)}=(a_{10},a_6,a_7,a_8,a_9)$ too. However, direct synthesis of $\pi_1^{(1)}$ and $\pi_2^{(1)}$ may be better or worse that synthesizing $\pi_1^{(2)}$ and $\pi_2^{(2)}$. To remove the effect of element ordering on the synthesis cost, we synthesize each two disjoint cycles in all possible ways (e.g., for two disjoint 5-cycles, 25 different ways are explored). Next, the best possible synthesis cost is assigned as the weight of the related edge.

\label{timecomplexity}
Assume that a specification with $k$ cycles of length $n_1$, $n_2$, ..., $n_k$ is given. A cycle of length $n_i$ ($1 \leq i \leq k$) can be decomposed in $N_{DCM}(n_i)$ different ways each of which includes $\lfloor n_i/5 \rfloor$ disjoint 5-cycles. Therefore, by selecting one of the available decompositions for each input cycle, $\sum_{i=1}^{i=k}{\lfloor n_i/5 \rfloor}$ disjoint 5-cycles will be resulted (Fig. \ref{Fig:CA1}) which lead to a complete graph with $\sum_{i=1}^{i=k}{\lfloor n_i/5 \rfloor} \times (\sum_{i=1}^{i=k}{\lfloor n_i/5 \rfloor}-1)/2$ edges (Fig. \ref{Fig:CA2}). Hence, the total time complexity required to select an appropriate cycle assignment for such decomposition is $25 \times \sum_{i=1}^{i=k}{\lfloor n_i/5 \rfloor} \times (\sum_{i=1}^{i=k}{\lfloor n_i/5 \rfloor}-1)/2 \times O(synthesis) + O(matching)$. Consideration of all possible cycle decompositions leads to $\prod_{i=1}^{i=k}{N_{DCM}(n_i)} \times (25 \times \sum_{i=1}^{i=k}{\lfloor n_i/5 \rfloor} \times (\sum_{i=1}^{i=k}{\lfloor n_i/5 \rfloor}-1)/2 \times O(synthesis) + O(matching))$. As can be seen, the time complexity of evaluating all possible cycle decompositions is very large. In the experimental results section, the runtime for each benchmark was limited to a reasonable time. Since no cycle decomposition is required for other elementary cycles, much less time will be required to select cycle pairs among the available 2-, 3- and 4-cycles.

\subsection{Post-Synthesis Optimization} \label {sec:post_opt}
Finding the optimal realization for a given reversible specification needs the evaluation of an exponential search space\footnote{Consider a quantum circuit of size $n$. Suppose that the optimal realization of a reversible specification needs $h$ gates from a library of size $M$. It can be verified that an exhaustive method needs the evaluation of $M^h$ gates where $M=$ O$(n\times 2^n)$ as follows:\\
There are C$^{1}_{n}$ possible NOT gates and C$^{2}_{n}$ possible CNOT gates in which one of its two inputs can be the target output. Hence, the total number of 2$\times$C$^{2}_{n}$ CNOT gates can be obtained. In contrast, for a ($k$+1)-bit gate, $k \in (2, 3,\cdots, n-1)$, there are $C^k_{n-1}$ possible gates when the target can be the $i^{th}$ ($i \in [1,n]$) bit. Considering all possible bits as the target leads to the total number of $n \times C^k_{n-1}$ ($k$+1)-bit gates. Therefore, the total number of gates is $C^1_n + 2\times C^2_n + n \times (\sum_{i \in (2\cdots n-1)} C^i_{n-1})= n \times 2 ^{n-1}$.}.
Therefore, it is very time-consuming to obtain an optimal realization for a given middle size reversible specification\footnote{The evaluation of synthesized circuits should be done with respect to a specific metric. Quantum cost or gate count can be used for this purpose.}. As a result, the usefulness of exact synthesis methods limits to relatively small specification. In addition, there are various metrics besides gate count or quantum cost \cite{Saeedi2008} that can be considered in the synthesis stage to improve the synthesized results. Altogether, due to various complexities involved in the synthesis of reversible circuits, there is a need to improve the quality of synthesized circuits in a post-processing step.

Previously, a few post-synthesis optimization methods have been introduced which used some pre-defined gate patterns (called \emph{templates}) \cite{Maslov2008} or a well-developed data structure \cite{Prasad2006} for the optimization of synthesized circuits.
In this paper, we use the method of \cite{Prasad2006} as a post-synthesis optimization algorithm as discussed in Section \ref{exp_results}.

\section {Experimental Results} \label {exp_results}
The proposed library-based synthesis methodology was implemented in C++ and all of the experiments were done on an Intel Pentium IV 2.2GHz computer with 2GB memory. In order to find a perfect matching on a given graph, we used Blossom V implementation \cite{Kolmogorov2008}. In addition, we used two recent synthesis tools proposed in \cite{Maslov2007a} and \cite{Prasad2006} for our comparisons. To the best of our knowledge, these are the most recent relevant works on reversible synthesis algorithms. In particular, \cite{Prasad2006} is similar to our synthesis algorithm with respect to using NCT gates and cycles. The application of exact methods like \cite{Grosse2009, Grosse2008} for finding optimal circuits are limited to small functions.

In all experiments, the post-synthesis optimization algorithm proposed in \cite {Prasad2006} was applied to simplify circuits produced by our synthesis methodology. In addition, the synthesis algorithm of \cite{Maslov2007a} was applied in `\emph{synthesized/ resynthesized using 3 methods}' mode for circuits with $n<15$ ($n$ is the circuit size) and in `\emph{synth/resynth with MMD (15+ variables)}' for $n > 15$. For \cite{Maslov2007a}, the synthesis algorithm, the templates matching method, the random and exhaustive driver algorithms were applied sequentially to synthesize each function with a time limit of 12 hours as in \cite{Maslov2007a}. Bidirectional and quantum cost reduction modes were also applied.

To evaluate the proposed synthesis methodology, the completely specified reversible benchmark functions (no DC) with more than six variables \cite{Grosse2008} were examined as library elements were designed for more than six variables in \cite{Sasanian2009}. Note that for small circuits, several well-developed exact and heuristic methods have been proposed \cite{Maslov2007a, Hung2006, Grosse2008, Gupta2006, Grosse2009}. We first fixed zero and 2$^i$ terms by applying a few Toffoli and CNOT gates in a pre-synthesis optimization step. Then, other parts of the proposed methodology were applied. To compare the results, we evaluated all synthesis algorithms in terms of quantum cost and the number of garbage bits. Quantum costs were calculated based on \cite{Maslov2008}.

The results of our synthesis algorithm and the previous best-proposed circuits that used the same gate library are reported in Table \ref{table:exper}. Headings `w/ g' and `w/o g' stand for `with garbage' and `without garbage', respectively. In addition, `\# g' denotes the number of garbage line. The symbol `-' is used if the algorithm fails to synthesize the circuit in 12 hours.

The synthesis tool of \cite{Maslov2007a} failed to synthesize the functions urf4 and urf6 after 12 hours. For urf1, urf2, urf3, and urf5 functions, several circuits were reported in \cite{Wille2008}. The resulted costs for these circuits are 45855, 16152, 121716, and 24253, respectively. Since applying the method of \cite {Maslov2007a} significantly improves the previous costs, we reported the new ones in Table \ref {table:exper}.

Since the number of valid decompositions for each cycle grows rapidly with the size of functions, for each benchmark function, we limited the runtime to 30 minutes and evaluated a limited set of decompositions for each cycle to find the best possible cost. Table \ref{table:additional_exper} shows the CPU time and the peak memory usage of the proposed synthesis methodology for each function. As illustrated in this table, the required CPU time for the decomposition step is less than five minutes for each circuit. In addition, the post-synthesis optimization step needs less than 5 minutes on average. The cycle-assignment step which includes the evolution of all possible cycle pairs for finding the best synthesis cost is the only time-consuming step. The required run time for other steps of Fig. \ref{Fig:flow} is negligible. As discussed, the best available synthesis algorithm needs about 12 hours to synthesize the available benchmarks (e.g., hwb11). Hence, the potential of the proposed synthesis methodology in synthesizing large function is considerable.

As demonstrated in Table \ref{table:additional_exper}, the proposed synthesis methodology needs up to 1.3 GBytes of memory to synthesize each benchmark function. In this table, the percentage of modified rows in truth-table representation was also reported. As discussed in Section \ref{sec:IF}, while all rows should be kept in memory for truth table representation, only modified rows need to be represented in CCF. As shown in Table \ref{table:additional_exper}, while for some functions (e.g., hwb11), the CCF representation is not very efficient compared with the truth-table representation, for some others (e.g., urf6) the CCF representation is very efficient. Altogether, CCF needs to represent about 20\% less rows on average.\label {CCFIF}

Table \ref{table:exper} shows the synthesis results. In this table, the synthesis cost of applying the method of \cite {Sasanian2009} for only one decomposition and with a trivial cycle assignment, where consecutive cycles are assigned to each other, are shown (1-Way DCM+CA). As shown in Table \ref{table:exper}, our synthesis costs for almost all functions are better than the costs of other methods. Since all of the attempted functions are even permutations, they can be implemented by the NCT-library with no additional garbage line \cite{Shende2003}. As the synthesis algorithm of \cite{Maslov2007a} uses one additional garbage line for the circuits of Table \ref{table:exper} (except ham7 and cycle10\_2) the synthesis costs with and without garbage lines are reported.

\begin{table}[!tb]
\caption{The CPU time and peak memory usage of the proposed synthesis methodology}
\label{table:additional_exper}
\centering
\scriptsize

\begin{tabular}{|c|c|c|ccc|c|}
\hline
 Benchmark & n & Modified rows  & \multicolumn{3}{c|}{CPU Time} & Peak memory \\
 Function & & (\%) & Decomposition & CA & Optimization & (Mbytes)\\
 & & & (Milliseconds) & (Minutes) & (Minutes) &  \\
\hline
ham7 			& 7 & 62 	&  3 		& 29 	& 1 	& 86 \\
\hline
hwb7 			& 7 & 92 	&  5 		& 29 	& 1 	& 87.5 \\
\hline
hwb8 			& 8 & 96 	&  5 		& 28	& 2 	& 90 \\
\hline
hwb9 			& 9 & 97 	&  14	 	& 25 	& 5 	& 99.5 \\
\hline
hwb10 		& 10 & 97 	&  18 		& 26 	& 4 	& 250 \\
\hline
hwb11 		& 11 & 99 	&  103 	  & 24 	& 6 & 263 \\
\hline
cycle10\_2 	& 12 & 43 	&  100600 & 23	& 5	& 980 \\
\hline
urf1 			& 9 & 92 	&  217011 & 21 	& 5	&  453 \\
\hline
urf2 			& 8 & 89 	&  43867 	& 26 	& 3 	& 290 \\
\hline
urf3 			& 10 & 93 	&  256352		& 20 	& 5	& 1220 \\
\hline
urf4 			& 11 & 98 	&  2568 	& 29 	& 1 	& 890 \\
\hline
urf5 			& 9 & 85 	&  711 	& 29 	& 1 	& 90 \\
\hline
urf6 			& 15 & $\sim{0}$ 	&  945 	& 29 	& 1 	& 530 \\
\hline
\hline
\end{tabular}
\end{table}

\begin{table}[!tb]
\caption{The comparison costs of our library-based synthesis methodology with the algorithms of \cite{Maslov2007a}, \cite{Sasanian2009} and \cite{Prasad2006}. Improved results are in bold both for {w/} and {w/o} garbage. For \cite{Maslov2007a}, a time limit of 12 hours was applied as done in \cite{Maslov2007a}. The method of \cite{Prasad2006} and \cite {Sasanian2009} required a few minutes for each function. At most 30 minutes were required for each circuit in the proposed methodology as shown in Table \ref{table:additional_exper}.}
\label{table:exper}
\centering
\small
\begin{tabular}{|c|ccc|c|c|c|}
\hline
 Benchmark & \multicolumn{3}{c|}{\cite{Maslov2007a} } & \cite{Prasad2006}& \cite {Sasanian2009} & Our method\\
 Functions   & \#g & w/ g & w/o g & (w/o g)  & (w/o g) & (w/o g) \\
\hline
\hline
ham7  		&0 & \textbf{49}	& \textbf{49} & 2695 & 2117 & 1804 \\
\hline
hwb7 			&  1	& \textbf{2609}	& \textbf{2613} & 4450 & 3177& 2727  \\
\hline
hwb8 			& 1	& \textbf{6197}	& 7015 & 10727 & 7163 & \textbf{6535}  \\
\hline
hwb9 			& 1	& 20378	& 22510 & 28135 & 16283 & \textbf{15462}  \\
\hline
hwb10 		& 1	& 46597	& 59197 &64442 & 36182 & \textbf{34224}  \\
\hline
hwb11 		& 1	& 122144	& 136760 &179966 & 91973 & \textbf{86942}  \\
\hline
cycle10\_2 	& 0	& \textbf{1206}	& \textbf{1206} & 197041 & 93086 & 89192  \\
\hline
urf1 			& 1	& 21850	& 23983 &31155 & 17281 & \textbf{16619}  \\
\hline
urf2 			&1	& 8161	& 9418 & 12823 & 7291 & \textbf{6600}  \\
\hline
urf3 			& 1	& 49843	& 61046 & 76114& 38133 & \textbf{36927}  \\
\hline
urf4 			& -	& -	& - & 190058 & 93992 & \textbf{90696}  \\
\hline
urf5 			& 1	& 12782	& 14225 &24086 & 14876 & \textbf{13930}  \\
\hline
urf6 			& -	& -	& -&34431 & 17367 & \textbf{16687}  \\
\hline
\end{tabular}
\end{table}

\section {Conclusion} \label {conc}
In this paper, a synthesis methodology for reversible circuits was proposed which used a set of building blocks and a library to synthesize a given specification. To this end, each input specification is considered as a permutation with several cycles where each cycle is synthesized by some reversible gates. If a given cycle is found in the library, it is synthesized directly; otherwise, the proposed decomposition algorithm detaches the building blocks from the given cycle. The decomposition algorithm explores all possible minimal and inequivalent factorizations where the number of disjoint cycles is maximized. To synthesize a given permutation, cycle pairs should be selected to reduce synthesis cost. Therefore, a cycle assignment algorithm was proposed based on the graph perfect matching algorithm too. Experimental results on reversible functions shows the advantage of the proposed approach in reducing both synthesis cost (i.e. quantum cost and number of garbage lines) and runtime.

\section*{Acknowledgment}
We would like to acknowledge Dmitri Maslov from University of Waterloo for providing an executable version of his synthesis tool.


\end{document}